\documentclass[journal,letterpaper,twoside]{IEEEtran_nssmic}
\usepackage{graphics}
\usepackage{graphicx}
\def\hbabar{\mbox{{\huge\bf\sl B}\hspace{-0.1em}{\LARGE\bf\sl A}\hspace{-0.03em}{\huge\bf\sl B}\hspace{-0.1em}{\LARGE\bf\sl A\hspace{-0.03em}R}}}

\def\babar{\mbox{\sl B\hspace{-0.4em} {\small\sl A}\hspace{-0.37em}
    \sl B\hspace{-0.4em} {\small\sl A\hspace{-0.02em}R}}}
\def\sbabar{\mbox{\sl B\hspace{-0.4em} {\scriptsize\sl
      A}\hspace{-0.37em} \sl B\hspace{-0.4em} {\scriptsize\sl
      A\hspace{-0.02em}R}}}

\def\CP      {\ensuremath{C\!P}}
\def\Bbar    {\kern 0.18em\overline{\kern -0.18em B}{}}
\def\BB      {\ensuremath{B\Bbar}} 
\def\cms     {\ensuremath{{\rm \,cm}^{-2} {\rm s}^{-1}}}

\begin{document}
  %
  \title{Measurement of the Charge Collection Efficiency after Heavy
    Non-Uniform Irradiation in \hbabar\ Silicon Detectors}
 
  %
  %
  \author{
    S. Bettarini, M. Bondioli, L. Bosisio, G. Calderini,
    C. Campagnari, S. Dittongo, F. Forti, M. A. Giorgi,
    G. Marchiori$^*$, G. Rizzo 
    \thanks{
      S. Bettarini, M. Bondioli, G. Calderini, F. Forti, M. A. Giorgi,
      G. Marchiori, G. Rizzo are with INFN-Pisa and Universit\`a 
      degli studi di Pisa.
    }
    \thanks{
      L. Bosisio and S. Dittongo are with INFN-Trieste and Universit\`a
      degli studi di Trieste.
    }
    \thanks{
      C. Campagnari is with University of California at Santa Barbara.
    }
    \thanks{
      M. A. Giorgi is also with Stanford Linear Accelerator Center.
    }
    \thanks{
      *Corresponding author. {\em E-mail address:} giovanni.marchiori@pi.infn.it
    }
  }
\maketitle

\begin{abstract}
We have investigated the depletion voltage changes, leakage
current increase and charge collection efficiency of a silicon
microstrip detector identical to those used in the inner layers of the
\babar\ Silicon Vertex Tracker (SVT) after heavy non-uniform
irradiation.
A full SVT module with the front-end electronics connected has been
irradiated with a 0.9 GeV electron beam up to a peak fluence of
3.5$\times$10$^{\bf 14}$ $e^{-}$/cm$^2$, well beyond the level causing
substrate type inversion.
We have irradiated the silicon with a non-uniform profile having
$\sigma$=1.4 mm that simulates the conditions 
encountered in the \babar\ experiment by the modules intersecting the
horizontal machine plane. The position dependence of the charge
collection properties and the depletion voltage have been investigated
in detail using a 1060 nm LED and an innovative measuring technique
based only on the digital output of the chip.  
\end{abstract}


\section{Introduction}

\PARstart{S}{everal} tests have been performed in the past to study
the effects of radiation damage to the \babar\
Silicon Vertex Tracker (SVT) sensors and to their front-end
electronics~\cite{bib:svt_radhard_studies}, but the reduction of
charge collection efficiency (CCE) 
after irradiation has never been measured quantitatively.
In addition, it has never been directly demonstrated that
a SVT module
can be operated normally after substrate
type-inversion. This is expected to happen at a dose of
$(3\pm 1)$ Mrad, based on measurements
performed on test structures from the same wafers.
To address these issues a module identical to those used in the inner
layer of the SVT, with the front-end electronics connected, has been
irradiated with a 0.9 GeV electron beam. A total peak fluence of
$3.5 \times 10^{\rm 14} e^{-}/{\rm cm}^2$, corresponding to a peak dose
of 9.3 Mrad, has been delivered to the silicon.
A second module, which has not been irradiated, has been used as a
control sample to track variations in the environmental
conditions.

\section{Radiation damage of the SVT}
The \babar\ experiment~\cite{bib:babar_detector}, at the 
SLAC PEP-II $e^+e^-$ storage ring~\cite{bib:pep-II},
has the primary physics goal of precisely measuring \CP-violating
asymmetries and rare branching fractions in $B$ meson decays.
\BB\ pairs are produced in head-on
collisions between 9.0 GeV electrons and 3.1 GeV positrons.
Since a very large sample of $B$ decays is needed, PEP-II was designed
to deliver the high peak luminosity of $3\times 10^{33} \cms$ (the
production cross section is $\sigma_{e^+e^-\to\BB}$ = 1.1 nb)

The Silicon Vertex Tracker~\cite{bib:svt} was installed in
\babar\ in early 1999 and has been reliably operated for five years,
providing excellent and efficient vertexing and tracking information.
It is composed of five layers of 300 $\mu$m thick, double-sided
microstrip detectors. p$^+$ strips on one side and
orthogonally-oriented n$^+$ strips on the other side
are implanted on a high-resistivity n$^-$ bulk. They are
AC-coupled to the electronics via integrated decoupling capacitors.
The detectors are operated in reverse mode at full depletion, with
bias voltage $V_{\rm bias}$ typically 10 V higher than the depletion
voltage $V_{\rm depl}$ (which lies in the range 25 V -- 35 V). The leakage
current per unit area under these conditions, prior to installation in
\babar, is lower than 100 nA/cm$^{2}$ at room temperature.

The main source of background in the SVT comes from electromagnetic
showers originating in the material of the beam-line by off-momentum
beam particles ($e^\pm$), which 
are over-bent by the permanent dipole magnets located in the proximity
of the interaction point to separate the two beams.
A significant fraction of this background is expected
to be composed of electrons and positrons with energies of a few
hundreds of MeV. Their non-ionizing energy loss (NIEL), using the
asymptotic value of the displacement cross section for 200 MeV
calculated in~\cite{bib:Summers}, is about twelve times lower than
that of 1 MeV neutrons at equal fluence, and the NIEL-normalized
damage constant $\alpha$/NIEL has been estimated in previous
measurements to 
be about one third than that of neutrons~\cite{bib:svt_radhard_studies}:
\begin{eqnarray}
{\rm NIEL (900 MeV }\ e^- {\rm )/NIEL(1 MeV}\ n) = 8.106\times 10^{-2}
\\
\frac{\alpha {\rm (900 MeV }\ e^- )}{{\rm NIEL (900 MeV }\ e^-) }
  \approx 1/3  \frac{\alpha{\rm (1 MeV}\ n)}{{\rm NIEL(1 MeV}\ n)}
\end{eqnarray}

The dose absorbed by the silicon, which is measured by means of 12 
silicon p-i-n diodes close to the inner layer of the SVT,
varies strongly as a function of the 
azimuthal angle around the beamline, and is highly peaked in a
narrow region of the (horizontal) bend plane of the machine, following
a roughly gaussian distribution with $\sigma \approx 2$ mm.
The inner layer of the SVT, located at a radius of 3.3 cm
from the beam line, receives the highest dose.
At design luminosity the average dose for the silicon of the inner
layer was expected to be 33 krad/yr, peaking to 240 krad/yr in
the horizontal region. The detectors were therefore originally designed to
withstand up to 2 Mrad of total radiation dose, which would have been
reached in ten years of running only in the inner horizontal
region of the SVT, and which is expected to be less than the
dose at which bulk type-inversion occurs.
However, excellent PEP-II performance has been 
significantly higher than expected. The peak instantaneous
luminosity has reached $9\times 10^{33} \cms$, three times
the design value, and is expected to increase up to $2\times 10^{34}
\cms$.
A thin horizontal region of the inner part of the detector has thus
already received the 
dose budget of 2 Mrad and will receive 9 Mrad by 2009, whereas a
larger fraction of silicon in the inner layers 
away from the horizontal plane should accumulate a
dose between 2 and 5 Mrad by the same date.

\section{Principle of the CCE measurement}
The basic idea of the measurement is to use a 1060 nm LED, whose
attenuation length $\lambda_{\rm att} = 1$ mm is deeper than the
300 $\mu$m of the silicon thickness, to generate charges in the
sensors. The charges then drift in the fully depleted silicon and the
signals induced on the microstrip electrodes on the two sides of the
detectors are amplified by the front-end electronics.

\begin{figure}[!htb]
  \vspace{0.1in}
  \begin{center}
    \includegraphics[width=3.5in]{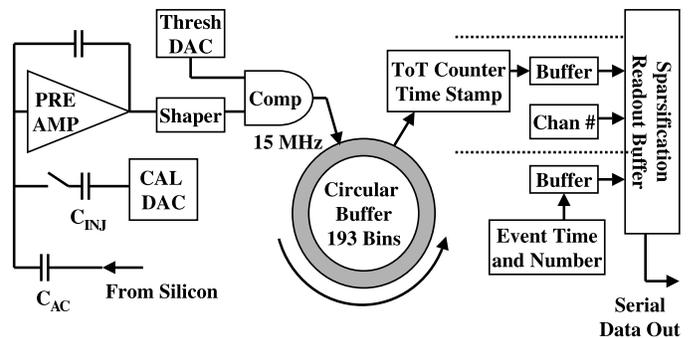}
    \caption{Schematic diagram of the AToM front-end IC.}
    \label{fig:atom_channel}
  \end{center}
\end{figure}

Each readout strip of the sensors, whose pitch is 50 $\mu$m on the
n-side and 100 $\mu$m on the p-side, is connected
to one of the 128 channels of the AToM IC~\cite{bib:ATOM}. The AToM IC
is a custom readout chip produced with a Honeywell rad-hard 0.8 $\mu$m CMOS
process. It is capable of simultaneous acquisition, digitization and
readout. 
Each channel of the AToM IC, as shown in
Figure~\ref{fig:atom_channel}, has an analog section consisting
of a low-noise charge-sensitive pre-amplifier followed by a CR-(RC)$^2$
shaper, whose output is coupled differentially into a comparator.
The nominal gain of the pre-amplifier is 250 mV/fC.
The shaping time is programmable, with a minimum of 100 ns, up to 400
ns. In our measurements we used the 100 ns setting, which is also used
in the modules in the inner layers of the SVT.
The comparator threshold is controlled by an on-chip 6-bit DAC
(Thresh DAC) whose least significant bit (LSB) has a nominal value of
12.5 mV, corresponding to a charge of 0.05 fC at the pre-amplifier's input.
The comparator output goes to an SRAM pipeline, which provides a trigger
latency of 12.93 $\mu$s. When the input to the comparator exceeds the
pre-set threshold, the output goes high and a series of
ones is clocked into the pipeline.
A calibration charge can be injected into
the preamplifier by means of a 50 fF capacitor (C$_{\rm inj}$) connected with a
switch to the pre-amplifier input and controlled by a 6-bit DAC
(CAL DAC). The CAL DAC LSB has a nominal value of 10 mV, corresponding
to a charge on C$_{\rm inj}$ of 0.5 fC.

Upon receipt of an external trigger, a
selectable region in the pipeline is searched for a zero-to-one
transition. The transition indicates a ``hit'' and the number of following ones
at the comparator output, divided by the clock period, represents the
time-over-threshold (TOT) of the hit, which is stored as a 4-bit
number and has an approximately logarithmic dependence on the input
charge. This allows charge measurement over a broad dynamic range
($\approx$ 40 fC) with a limited number of bits, but the limited accuracy
makes the TOT unsuitable for the purpose of our measurement, where a
good analog resolution is needed in order to establish small drops in
the CCE. Therefore we have turned to an alternative method of
measuring the charge, based on ``threshold scans''~\cite{bib:campagnari}.

\begin{figure}[!htb]
  \begin{center}
    \includegraphics[width=2.5in]{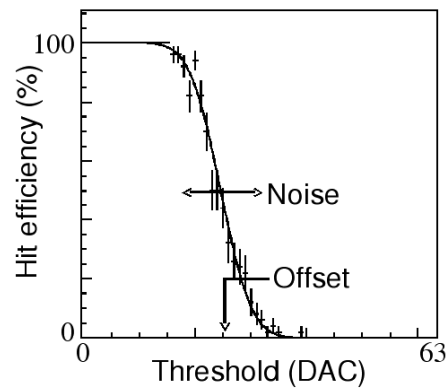}
    \caption{Schematic representation of a threshold scan for a single
      readout channel.}
    \label{fig:tscan}
  \end{center}
\end{figure}

A threshold scan consists in measuring, for each read-out channel $i$
at fixed injected charge $Q_i$ at the input of the pre-amplifier, the
hit efficiency as a function of the pre-set threshold of the
comparator, which is varied inside the full dynamic range.
The 50\% turning point of the hit efficiency versus threshold 
distribution is the threshold offset $V_{\rm off}(i,Q_i)$ for the channel
$i$ at charge $Q_i$, as shown in Figure~\ref{fig:tscan}.
The threshold offset has a linear dependence on the charge which
is given by the following expression:
\begin{equation}
  V_{\rm off}(i,Q_i) = P_i + Q_i \times G_i
\end{equation}
The pedestal $P_i \equiv V_{\rm off}(i,Q_i=0)$ and the gain $G_i
\equiv dV_{\rm off}(i,Q_i)/dQ_i$ of the $i$-th electronic 
channel can be accurately determined by means of calibration
threshold scans, in which the silicon is not illuminated by the LED,
and a known amount of charge $Q_i$ is injected at the input of the
pre-amplifier by closing the switch which connects it to the 50 fF capacitor
C$_{\rm inj}$.
$P_i$ and $G_i$ are found by performing calibration threshold scans with
different known charges $Q_i$ and fitting the $V_{\rm off}$ vs $Q_i$
distribution with a linear function.

When the LED is used to generate charge in the silicon, the charge at
the input of the pre-amplifier of the $i$-th channel is $Q_i =
Q_{\rm LED}\times f(i) \times {\rm CCE}_i$, where $Q_{\rm LED}$ is
the total charge release by the LED in the silicon, which 
is proportional to the LED intensity, ${\rm CCE}_i$ is the local charge
collection efficiency of the detector and $f(i)$ accounts for division
of the charge between the channels and the angular distribution of the
LED power. When the detector is fully depleted and the electric field
intensity is sufficient to collect almost all the charge, $\Sigma_i
f(i)=1$ (for both sides) and the CCE, as a function of the reverse bias
voltage applied to the silicon, saturates. 
In our case the LED is current-driven and the light flux is  
proportional to the LED current, therefore $Q_{\rm LED} = a \times I_{\rm 
  LED}$. In a threshold scan performed with the silicon illuminated by
the LED with current $I_{\rm LED}$ therefore the offset of the $i$-th
channel is given by
\begin{equation}
  V_{\rm off}(i,I_{\rm LED}) = P_i +  a \times I_{\rm LED} \times f(i)
  \times {\rm CCE}_i \times G_i 
\end{equation}

From threshold scans at different values of the LED current
one can extract for each channel the slope
$S_i \equiv a \times f(i) \times G_i \times {\rm CCE}_i$ by performing
a linear fit to the $V_{\rm off}(i,I_{\rm LED})$ vs $I_{\rm LED}$
distribution. 
By dividing the slope $S_i$ for the electronics gain $G_i$ and summing over
all channels we obtain therefore a quantity which is proportional to
the average CCE in the silicon region illuminated by the LED:
\begin{equation}
a \times \langle {\rm CCE} \rangle = \Sigma_i S_i/G_i
\end{equation}

By comparing the sum $\Sigma_i S_i/G_i$ before and after the irradiation
of the detectors we can therefore monitor the relative CCE drop.

\section{Experimental setup}

\subsection{Setup for the CCE measurement}
\begin{figure}[!htb]
  \begin{center}
    \includegraphics[width=3.5in]{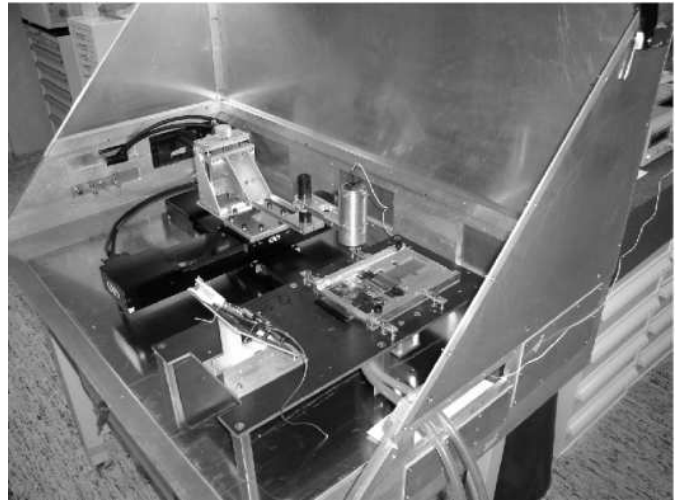}
    \caption{Mechanical setup for the CCE measurements. The silicon
      module is mounted in a metal frame, and above it the brass cylinder
      holding the LED is clearly visible. The LED is moved above the
      detector surface by means of the black X-Y stage.}
    \label{fig:mechanical_setup_CCE}
  \end{center}
\end{figure}

The LED is a current-driven EG\&G C30116 model,
with peak wavelength $\lambda_{\rm peak}$ = 1060 nm, risetime 
$t_{\rm rise} <$ 10 ns, typical peak flux vs LED current $\phi_{\rm
  peak}/I_{\rm LED} = 2$ mW/A.
It is connected through a 1 k$\Omega$ resistor to a 
GPIB-controlled pulser.
A thin brass foil with a 500 $\mu$m diameter
pinhole is placed at a distance of 3 mm from the LED lens surface.
A converging lens with focal length $f$ = 45 mm is placed at a distance of
90 mm from the pinhole and at the same distance from the module plane,
thus ensuring that the pinhole image is focused in the module plane.
In our measurements the light emitted from the LED enters the silicon
from the n$^+$-doped (ohmic) side.
The dimension of the luminous spot has been chosen to be
narrower than the region in which the CCE, after irradiation with a
beam with $\sigma \approx $ 2 mm, is supposed to change, but at the
same time it is large enough that the uncertainty in the relative alignment of
the LED and the silicon detectors ($<$100 $\mu$m) has a negligible impact
on the uncertainty on the measured CCE.
The LED, the pinhole and the lens are mounted inside a brass cylinder
attached to a mechanical arm of a GPIB-controlled X-Y stage.
A picture of the mechanical setup is shown in
Figure~\ref{fig:mechanical_setup_CCE}.

\begin{figure}[!htb]
  \begin{center}
    \includegraphics[width=3.5in]{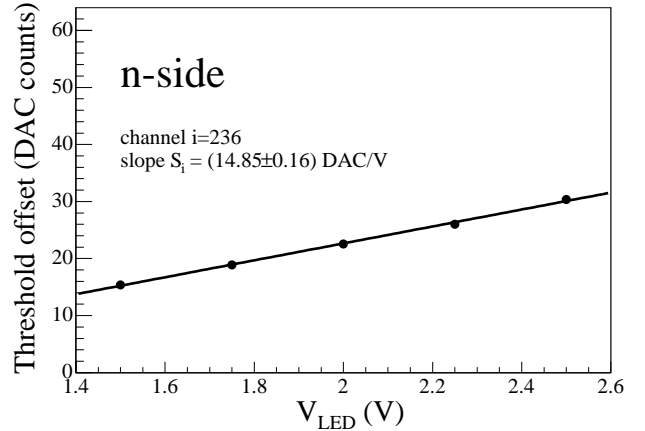}
    \caption{Distribution, for one readout channel illuminated by the LED,
      of the threshold offset $V_{\rm off}$ (the 50\% hit efficiency
      turn-on point) as a function of the amplitude of the pulse driving
      the LED (dots). The result of a linear fit to this distribution
      is superimposed (solid line).}
    \label{fig:linearity}
  \end{center}
\end{figure}

The charge generated in the silicon is controlled by
changing the amplitude of the pulse driving the LED. We have selected
a range of amplitudes in which the
LED response is linear and the signal at the shaper output remains
within the limited dynamic range of the THR-DAC, as shown in
Figure~\ref{fig:linearity}. 

\begin{figure}[!htb]
  \begin{center}
    \includegraphics[width=3.5in]{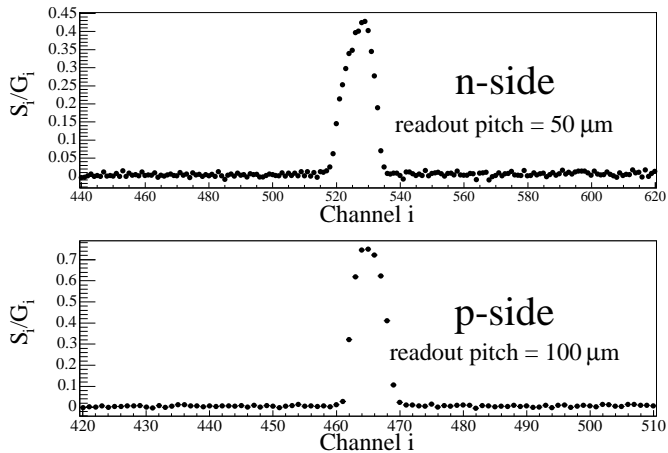}
    \caption{Distribution of the fitted slopes $S_i$ divided by the
      electronic gain $G_i$ as a function of the readout channel $i$. A peak
      in the channels illuminated by the LED is clearly visible. The
      integral of the peak is proportional to the 
      average charge collection efficiency in the illuminated area.
    }
    \label{fig:LED_spot}
  \end{center}
\end{figure}

The measurement process is fully automated: a workstation controls
the motion of the X-Y stage and the voltage setting of the pulser via
GPIB connections, while at the same time controls through an 
ethernet interface a VME-based computer which is responsible for
sending trigger signals to the pulser and to the readout section
of the front end electronics.
An example of the distribution of $S_i/G_i$ vs
channel measured with this setup is shown in
Figure~\ref{fig:LED_spot}: the peak width is consistent with
the pinhole size and the readout pitch of the strips.
Once the $S_i/G_i$ distribution has been measured, an offline
algorithm identifies the position of the peak, evaluates the baseline
in channels away from the peak, and computes the baseline-subtracted
sum $\Sigma_i S_i/G_i$ over the whole peak, which is proportional to
the average CCE in the point illuminated by the LED.
Repeated tests on the control module have shown that the
single measurement resolution is about 2\%. Alignment in the relative
position between the module and the X-Y stage and aplanarity effects
have been found to introduce negligible degradation in the CCE
resolution. Moreover, no significant variation (compared to the 2\%
intrinsic resolution) of the measured CCE has been observed over a
period of several hours of continuous operations.
However, differences up to 10\% have been found in CCE measurements
performed in different runs.
For this reason the CCE measured for the irradiated module has always 
been normalized to the average CCE measured, with a high number of
samplings, in a fixed set of points of the control module in
the same environmental conditions.


\subsection{Setup for the silicon irradiation}
\begin{figure}[!htb]
  \begin{center}
    \includegraphics[width=3.5in]{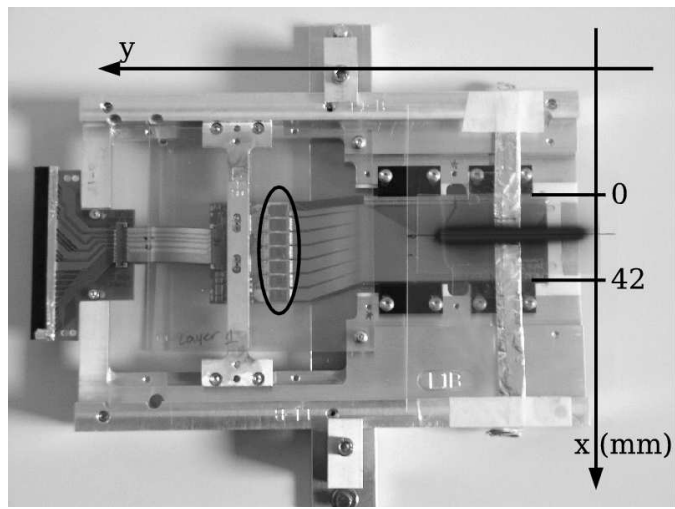}
    \caption{Picture of the irradiated module.
      During the irradiation the beam ($z$ axis) comes from the side not
      shown here.
      The radiochromic film keeps track (dark stripe at x$\approx$21 mm)
      of the beam position.
      The aluminum foil contains test structures used as a cross-check
      for the dose evaluation as described in the text.
      The ellipsis at the center surrounds the front-end
      electronics chips.
    }
    \label{fig:module_irradiated}
  \end{center}
\end{figure}

\begin{figure}[!htb]
  \begin{center}
    \includegraphics[width=3.5in]{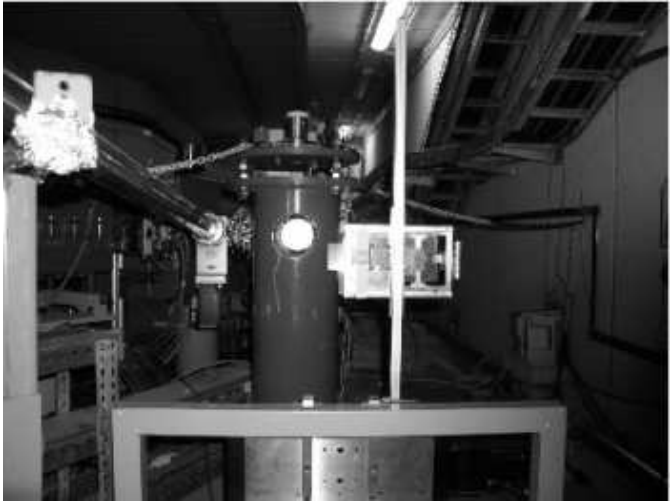}
    \caption{Mechanical setup of the module for the irradiation. The
      module (at the center) is mounted on a X-Y stage (bottom) in
      proximity of the end-flange of the Elettra linac.
    }
    \label{fig:setup_at_elettra}
  \end{center}
\end{figure}

The module irradiation has been performed at the Elettra Synchrotron
facility in Trieste with a 0.9 GeV electron beam.
The irradiation has been performed in six steps reaching a total peak
dose of about 9 Mrad.
Two pictures of the module setup are shown in
Figures~\ref{fig:module_irradiated} and~\ref{fig:setup_at_elettra}.

During each irradiation step the module is mounted
on a X-Y stage,
which is located at the end of the Linac and is remotely
controlled from the Linac Control Room through a serial connection.
The alignment of the module and of the X-Y stage is such that the
detector surface lies in a plane $xy$ orthogonal to the beamline ($z$
axis), and the module can be moved, by means of the X-Y stage, along
the $x$ and $y$ directions.
The silicon surface is a rectangle measuring $42.4$ mm in the $x$
direction, which is parallel to the orientation of the p$^+$ 
strips, and 82.6 mm in the $y$ direction, which is parallel to the
orientation of the n$^+$ strips.
The module thickness along the $z$ direction is 300 $\mu$m.

\begin{figure}[!htb]
  \begin{center}
    \includegraphics[width=2.0in]{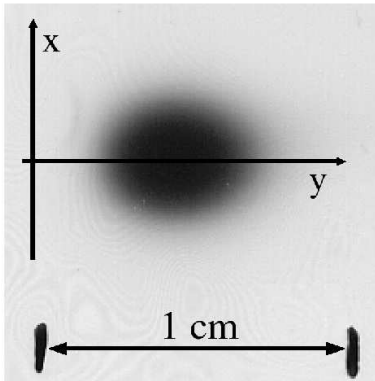}
    \caption{Beam spot obtained on a radiochromic dosimetry film
      irradiated for a few seconds. 
    }
    \label{fig:beam_spot_film}
  \end{center}
\end{figure}

Prior to the irradiation a radiochromic dosimetry film is placed
close to the surface of the silicon, to keep track of the beam position
during the irradiation. Another radiochromic film is placed above the
chip region to check that the electronics does not receive a
significant dose, thus avoiding the need in the CCE measurement to
disentangle effects caused by radiation damage to the silicon from
effects caused by damage to the electronics.
A third radiochromic film is attached to the edge of the metal frame
which hosts the module and is irradiated for a few seconds (to avoid
saturation) to obtain an image of the beam profile, which is necessary
to perform the alignment between the detector and the beam and is used
to estimate the beam profile. The beam spot obtained with 
this method is shown in Figure~\ref{fig:beam_spot_film}.
Two sets of test structures from the same wafer as the silicon 
detectors are mounted on the module frame in a position which corresponds 
to the center of the zone to be irradiated.


After alignment of the module with respect to the beamline, the linac
is turned on and the module is moved back and forth 
several times in the plane $xy$, along a
line parallel to the $y$ axis.
The beam has a gaussian profile whose width along the $x$ axis,
accurately measured by digitizing and analyzing the spot obtained on
the radiochromic dosimetry film irradiated for a few seconds, is
$\sigma_x \approx $1.4 mm.
By moving the module with constant velocity $v_y = $ 1 mm/s along $y$,
a 50 mm long region of the silicon is irradiated, with a dose profile
which is uniform in $y$ and gaussian in $x$, centered on the axis of
the module and with width equal to the beam width $\sigma_x$.
The beam remains several centimeters from the chips during all the
irradiation.

The peak fluence $\phi_e$ and hence the peak dose are determined from the beam
spread, the speed at which the module is moved during the irradiation,
the number of times $N_{\rm sweeps}$ this operation is performed and the
linac current $I_{\rm linac}$ through the relation
$\phi_e = N_{\rm sweeps}(I_{\rm linac}/q) / (\sqrt{2\pi} v_y \sigma_x)$ 
where $q = 1.6 \times 10^{-19}$ C is the electron charge.
For typical values ($\sigma_x$ = 1.4 mm, $v_y$ = 1 mm/s, $I_{\rm linac}$ = 30
nA and $N_{\rm sweeps}$ = 10), a total peak fluence of about $5.3 \times
10^{13} e^-$/cm$^2$ (corresponding to a peak dose of about 1.4 Mrad)
is delivered to the module in about ten minutes.
The linac current is measured with 4\% accuracy from the
current flowing in a toroidal coil coaxial with the beam
and is the dominating source of uncertainty in the peak fluence
estimate. A less precise dose estimate, obtained from the
increase in leakage current in the test structures irradiated with the
detector, is used as a cross check and is consistent with the estimate
from the linac current.

\section{Results}

\subsection{Leakage current increase}
\begin{figure}[!htb]
  \begin{center}
    \includegraphics[width=3.5in]{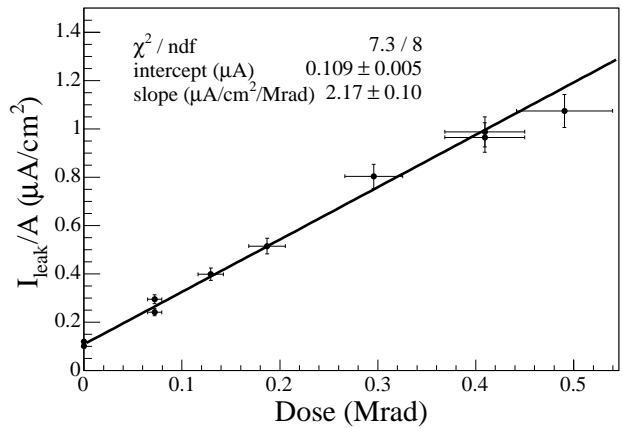}
    \caption{Leakage current increase per unit area after
      irradiation. The silicon area $A$ is 42.4x82.6 mm$^2$.}
    \label{fig:leakage_current_vs_dose}
  \end{center}
\end{figure}

After each irradiation step we have measured the leakage current of
the module from standard I-V curves at reverse bias.
The leakage current per unit area increases linearly with the
accumulated dose of 
the whole module, as shown in Figure~\ref{fig:leakage_current_vs_dose}.
The current increase per unit area vs dose is $(2.17\pm0.10)\ \mu{\rm
  A}/{\rm cm}^2/{\rm Mrad}$, normalized at a temperature of $23^\circ$C,
consistent with the increase observed in the SVT during \babar\ operation.
The damage constant $\alpha \equiv \Delta J_{\rm leak}/\langle\phi_e\rangle$,
where $J_{\rm leak}$ is the leakage current density and
$\langle\phi_e\rangle$ is the average electron fluence delivered to
the silicon, is $\alpha=(1.43\pm0.07)\times 10^{-18}$ A/cm,
normalized at $20^\circ$C. 

\subsection{Depletion voltage shift}
\begin{figure}[!htb]
  \begin{center}
    \includegraphics[width=3.5in]{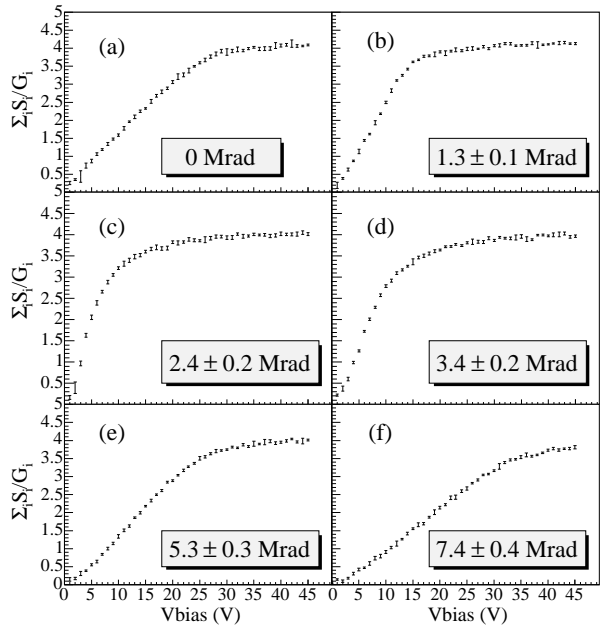}
    \caption{Measured $\Sigma_i S_i/G_i$ as a function of bias voltage,
      for the same point of the silicon and six different integrated
      doses. The plotted quantity is proportional to the 
      charge collection efficiency of the silicon.}
    \label{fig:CCE_vs_Vbias}
  \end{center}
\end{figure}

After each irradiation we have measured, at a point in the center of
the most irradiated zone of the detector, the sum $\Sigma_i S_i/G_i$ 
(which is proportional to the local CCE) as a function of the
reverse bias voltage $V_{\rm bias}$ applied to the silicon.
This is shown, up to the fifth irradiation step, in
Figure~\ref{fig:CCE_vs_Vbias}.

\begin{figure}[!htb]
  \begin{center}
    \includegraphics[width=3.5in]{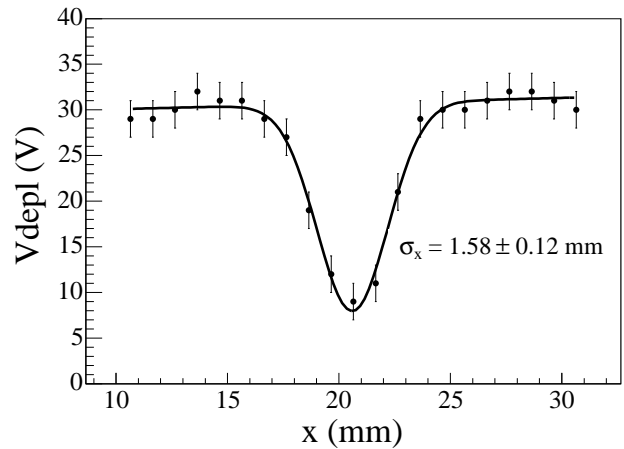}
    \caption{Depletion voltage as a function of the position, after two
      irradiation steps (total peak dose $\approx$ 2.5 Mrad at x=20.5 mm),
      for a set of points on a line orthogonal to the irradiation direction.}
    \label{fig:beam_profile_with_Vdepl}
  \end{center}
\end{figure}

For high values of the bias voltage the $\Sigma_i S_i/G_i$ vs $V_{\rm
  bias}$ curve is characterized by a plateau over which the CCE
increases very little when the bias voltage is increased,
corresponding to the fact that 
the trapping probability of charge carriers slowly decreases as long
as the electric field inside the detector becomes stronger.
We use the voltage value $V_{\rm sat}$ below which the sum $\Sigma_i
S_i/G_i$ and therefore the CCE fall below 90\% of the plateau value
as a rough estimator for the depletion voltage $V_{\rm depl}$ of the
silicon in the point illuminated by the LED.
As an example, consider the curve shown in 
Figure~\ref{fig:CCE_vs_Vbias} (a), measured before the first
irradiation step: in that case we estimate $V_{\rm 
  sat} \approx 30$ V, which is roughly consistent with the depletion
voltage $V_{\rm depl}=25$ V measured in structure tests from the same wafer. 
$V_{\rm sat}$ is not a good estimator for the depletion voltage
when $V_{\rm depl}$ is close to 0 V, because in that case there is a
bias voltage range in which the detector is fully depleted but the
electric field inside it is not intense enough to approach signal
saturation. In that case $V_{\rm sat}$ gives only an upper limit for
$V_{\rm depl}$: indeed, in all our measurements $V_{\rm sat}$ never
takes values below $\approx$ 10 V. However, when the depletion voltage
is higher than 10 V then $V_{\rm sat}$ is a reasonable estimator for $V_{\rm
  depl}$.
To confirm that $V_{\rm sat}$ is a reasonable estimator of the
depletion voltage, which is shifted by bulk damage of the silicon, we
estimate -- after the 
second irradiation step, corresponding to a total peak dose of 2.5
Mrad -- $V_{\rm sat}$ in a set of 21 points, equally spaced (at 1mm
steps), on a line which is orthogonal to the irradiation direction and
crosses the irradiated region. The distribution of $V_{\rm sat}$,
shown in Figure~\ref{fig:beam_profile_with_Vdepl}, exhibits a clear peak
whose width is consistent with the measured beam spread.

\begin{figure}[!htb]
  \begin{center}
    \includegraphics[width=3.5in]{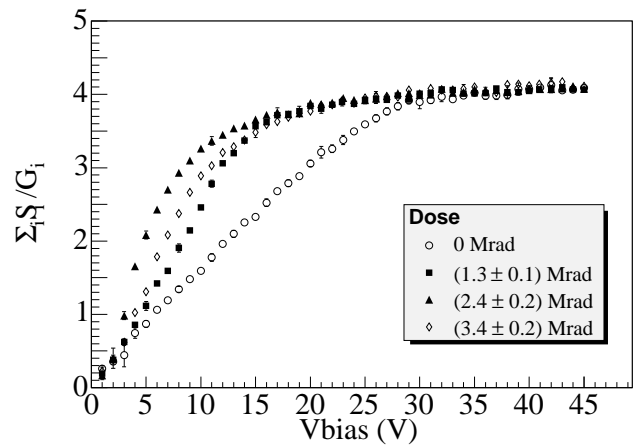}
    \caption{The curves of Figure~\ref{fig:CCE_vs_Vbias} (a), (b), (c) and
      (d) overlaid. The measured $\Sigma_i S_i/G_i$ is shown as a function
      of bias voltage, for the same point of the silicon prior to
      irradiation and after the first three irradiation steps.}
    \label{fig:CCE_vs_Vbias2}
  \end{center}
\end{figure}

From the measured values of $V_{\rm sat}$ at different doses obtained
  from the CCE vs $V_{\rm bias}$ curves of
  Figure~\ref{fig:CCE_vs_Vbias}, we see that
the depletion voltage in the damaged 
silicon first decreases with dose, then starts to increase again. 
This is evident when overlaying the curves in the same plot, as 
shown in Figure~\ref{fig:CCE_vs_Vbias2}. 
The inversion occurs between $(1.3\pm0.1)$ Mrad and $(3.4\pm0.2)$
Mrad: we therefore estimate the inversion point, which corresponds to
the bulk type inversion of the silicon, to be at $(2.4\pm1.0)$ Mrad.
After type inversion the detector continues to operate without
any problem.

\subsection{Charge collection efficiency drop}
\begin{figure}[!htb]
  \begin{center}
    \includegraphics[width=4.0in]{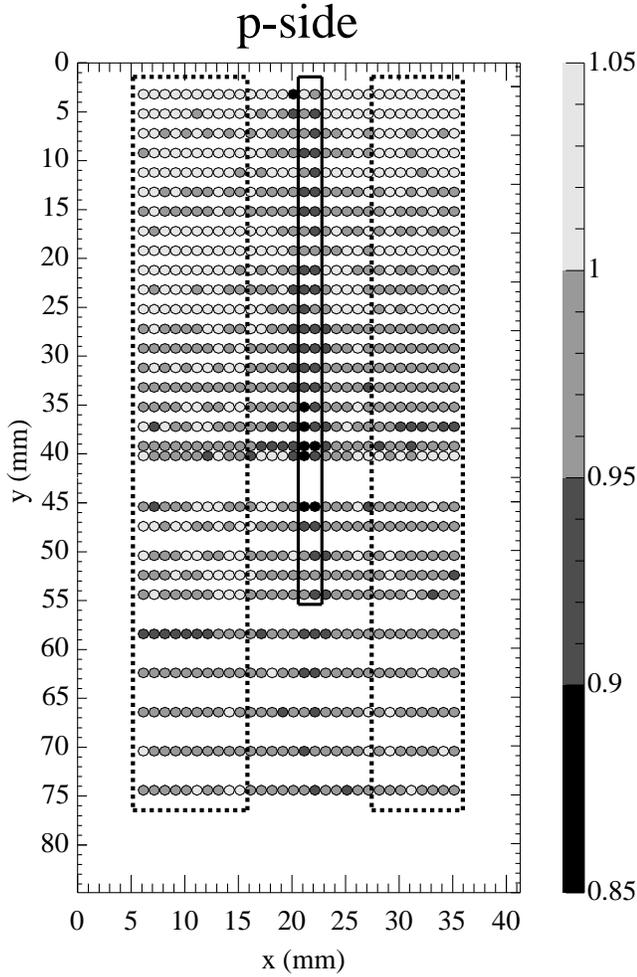}
    \caption{Ratio between the CCE after and before irradiation, as a
      function of the position on the module. Only 
      the p-side is shown.
      The central solid rectangle surrounds the
      most irradiated region (the average dose in the points enclosed by
      the rectangle is $(8.5 \pm 0.8)$ Mrad).
      The points inside the two lateral dashed rectangles are at least
      4$\sigma$ away from the beam axis and have received a negligible
      dose.}
    \label{fig:cce_pside}
  \end{center}
\end{figure}

The CCE has been measured, before and after irradiation, in a
grid of 30x30 points spanning almost the entire surface of the 
module.
These measurements have been performed with the detector reversed
biased with a potential $V_{\rm bias}$ chosen to be at least 10 V
higher than the estimated depletion voltage in the most irradiated
area of the silicon, to make sure that the silicon is fully depleted
everywhere across the detector.
The distribution of these points across the detector surface
and the ratio between the CCE measured after the last irradiation
step and the CCE before irradiation, for the p-side, is shown in
Figure~\ref{fig:cce_pside}.

\begin{figure}
  \begin{center}
    \includegraphics[width=3.5in]{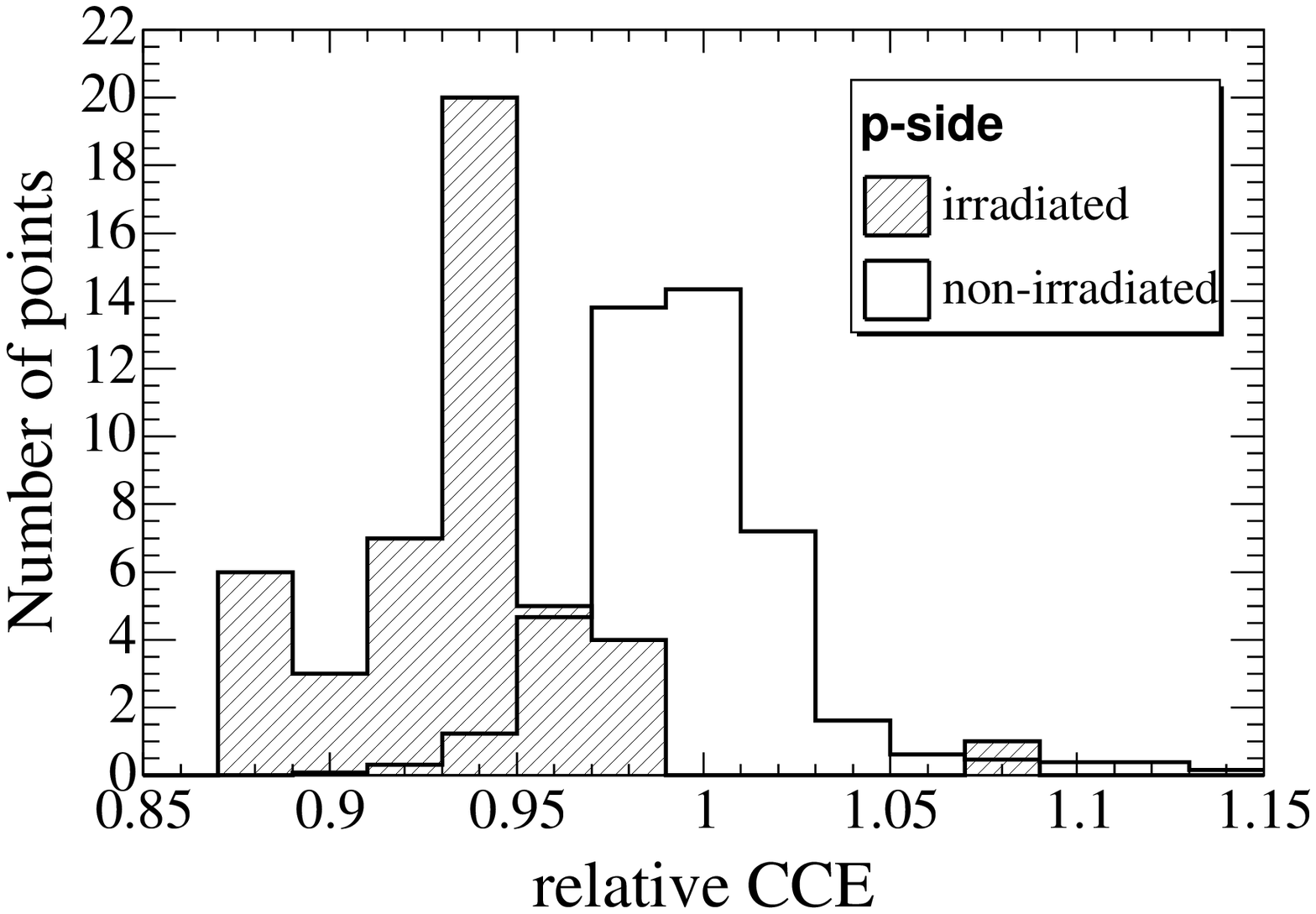}
    \includegraphics[width=3.5in]{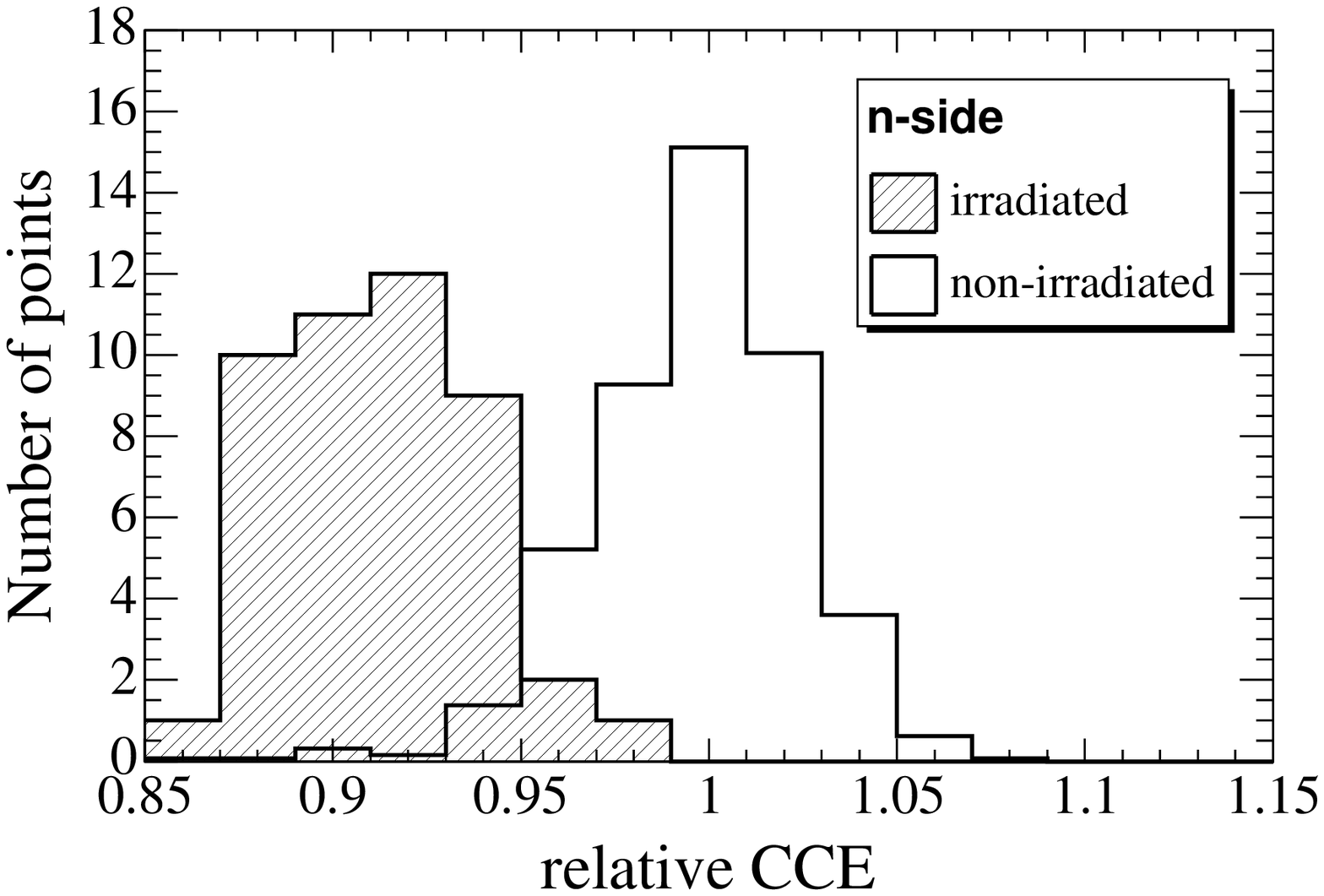}
    \caption{Ratio between the CCE after and before
      irradiation, for points which have received
      a dose of $(8.5 \pm 0.8)$ Mrad (hatched histograms) and points which
      have received a negligible dose (hollow histograms). Top:
      p-side. Bottom: n-side.}
    \label{fig:cce_irr_vs_nonirr_pside}
  \end{center}
\end{figure}

In Figure~\ref{fig:cce_irr_vs_nonirr_pside} we compare the CCE drop in
points at the center of the irradiated zone,
(the solid central rectangle in Figure~\ref{fig:cce_pside})
which have received a total dose of $(8.5\pm0.8)$ Mrad, with points which are 
at least 4$\sigma$ away from the irradiation axis and have received a
dose of only a few krad.
(the two lateral dashed rectangles in
Figure~\ref{fig:cce_pside}).
For points in the irradiated zone we measure a CCE decrease equal to
$(6\pm2)\%$ on the p-side and $(9\pm2)\%$ on the n-side, while no CCE
decrease is observed for points which have received a negligible dose.

\section{Conclusion}
We have irradiated, with a 0.9 GeV $e^-$ beam, a silicon detector
identical to those in the \babar\ Silicon Vertex Tracker. 
We have implemented a fully automated setup by means of which we can
generate charge in the silicon through illumination with
a 1060 nm LED and measure the charge collection efficiency through an
innovative method based on the digital output of the front-end
electronics connected to the detector. We have measured an increase in
leakage current per unit area at 23$^\circ$C of $(2.17\pm
0.10)\ \mu$A/cm$^2$/Mrad.
The detector, whose initial depletion voltage was 25 V, has undergone
type inversion at a dose of $(2.4\pm1.0)$ Mrad, after which it has
continued to operate without any problem. We have measured, in 
points irradiated with a dose of $(8.5\pm0.8)$ Mrad, a moderate charge
collection efficiency decrease equal to $(6\pm2)\%$ on the p-side and
$(9\pm2)\%$ on the n-side. We have thus demonstrated that the SVT
sensors can be operated for the whole lifetime of the \babar\
experiment and that bulk damage in the silicon will cause only a
modest impact on their performances.




\end{document}